\documentclass[a4paper,single column,12pt]{article}
\usepackage{epsfig,amsmath,amssymb,subfigure,placeins}
\usepackage{graphicx}% Include figure files
\usepackage{dcolumn}% Align table columns on decimal point
\usepackage{bm}% bold math
\usepackage{subfig}
\usepackage{float}
\usepackage{tabularx}
\usepackage{multirow}% http://ctan.org/pkg/multirow
\usepackage{hhline}% http://ctan.org/pkg/hhline
\usepackage{color}
\newcommand{\old}[1]{}

\newcommand{\be}{\begin{equation}}
\newcommand{\ee}{\end{equation}}
\newcommand{\ba}{\begin{eqnarray}}
\newcommand{\ea}{\end{eqnarray}}
\newcommand{\bi}{\begin{itemize}}
\newcommand{\ei}{\end{itemize}}

\begin{document}
\begin{flushright}
{\normalsize
%June 21 2008}
}
\end{flushright}
\vskip 0.1in
\begin{center}
{\large {\bf Quarkonium Dissociation  at Finite Chemical Potential}}
\end{center}
\vskip 0.1in
\begin{center}
Uttam Kakade\footnote{usk11dph@iitr.ac.in, uskakade@gmail.com}, 
Binoy Krishna Patra\footnote{binoyfph@iitr.ac.in}\\
%and Lata Thakur\footnote{lata1dph@iitr.ac.in}\\
{\small {\it Department of Physics, Indian Institute of
Technology Roorkee, India, 247 667} }
\end{center}
\vskip 0.01in
\addtolength{\baselineskip}{0.4\baselineskip} %wide line spacing
%opening
\section* {Abstract}
We have studied the dissociation of quarkonia states in a deconfined medium of quarks and gluons at the large baryon chemical potential and small temperature region. The aim of this study is to probe the dense baryonic medium expected to be produced at the  Facility for Anti-proton and Ion Research (FAIR), GSI Darmstadt. This is done by correcting both the short- and long-distance terms of the Cornell potential by a dielectric function, embodying the effects of deconfined quarks and gluons, at finite baryon chemical potential and temperature. It is found that $J/ \psi$ is dissociated approximately at 1.4$\mu_c$ in the temperature range 20-50 $MeV$, which can indirectly help to locate the point on the QCD phase diagram at the large chemical potential and low-temperature
zone.

\noindent PACS:~~ 12.39.-x,11.10.St,12.38.Mh,12.39.Pn
12.75.Nq, 12.38.Gc \\
\vspace{1mm}
\noindent{\bf Keywords}: heavy ion collision, QCD phase transition, QGP, nonperturbative effects, quarkonium dissociation \\

\section{Introduction}
Relativistic heavy-ion collisions provide the unique opportunity of 
creating a hot and dense nuclear matter in the laboratory. According 
to statistical quantum chromodynamics (QCD), nuclear matter may 
undergo a color deconfined partonic phase, the quark gluon plasma (QGP), 
at sufficiently high temperature and/or density. Over the past few
decades, strenuous efforts have been made to devise clean
and experimentally viable signals that can unambiguously
identify the existence of QCD phase transition and trace
out its signatures. Charmonium (a bound state of charm
and anti-charm quarks) suppression had been predicted as a
signature for the deconfinement transition for a long time~\cite{Matsui:PLB178/1986}, where the Debye screening of the colored 
partons was the dominant mechanism of suppression. Experimental 
investigations revealed a noticeable reduction of the charmonium
production in proton-nucleus ($p$-$A$) collisions compared to the
hadronic collisions (scaled with binary nucleon-nucleon ($p$-$p$)
collisions)~\cite{Alessandro:EJPhy39/2005,Blaizot:PRL85/2000,
Grandchamp:PRL92/2004,Arnaldi:PRL99/2004}. A thorough understanding of this normal
nuclear suppression is essential in order to establish a robust
baseline reference, with respect to the anomalous suppression
~\cite{Aberu:PLB410/1997,Alassandro:EPJC48/2006}, pertaining 
to the formation of a deconfined partonic medium observed at 
Relativistic Heavy Ion Collision (RHIC) 
and Large Hadron Collider (LHC) energies that mainly scan the high 
temperature and 
almost vanishing baryon density region of the QCD phase 
diagram~\cite{Dpal:EPJC17/2000, BKPDKS:PLB505/20011}.
 Depending upon the magnitude of energy deposited in
the center-of-mass frame of colliding nuclei, there are in
general two scenarios in heavy-ion collision. The first one
is possibility of the complete stopping of the colliding
nuclei if energy density $\epsilon \approx $ 5 GeV/$ {\rm{fm}}^3 $,
creating sufficiently
high baryon density (parton density) to produce QGP. Full
transparency of colliding nuclei, the other scenario, takes place
if $\epsilon \sim $ 100 GeV/${\rm{fm}}^3$ creating a favorable environment to
produce QGP by vacuum polarization and leading to vanishing
chemical potential. Exploration of high baryon densities and
the moderate-temperature region of the QCD phase diagram is possible with the upcoming compressed baryonic matter
(CBM) experiment at the Facility for Anti-proton and Ion
Research (FAIR). At FAIR, the light and heavy ions can be collided 
in the beam energy range $E_{Lab}$ = 10-40 $A GeV$ to create
an environment favorable to investigation of extremely dense
nuclear matter in the laboratory through the measurements of
bulk and rare probes. Model calculations based on transport
and hydrodynamical equations~\cite{Arsene:PRC75/2007} predict 
that the highest net baryon densities ($\rho_B$) produced in 
the center of collision is $\sim$ 6 to 12 times the density of normal 
nuclear matter for the most central collision (b=0).
 
The FAIR will open avenues to investigate some of
the fundamental yet enigmatic issues of strong interaction
thermodynamics at large baryon density. Some of the key
issues in this region are the study of hadronic properties
in dense nuclear matter, the deconfinement phase transition
from hadronic to quark-gluon matter driven by high baryon
densities, and the nuclear equation of state (EOS) at high
baryon densities. Thus the behavior of QCD at high baryon
density and moderate to low temperatures is interesting and has
potential applications to cosmology, astrophysics of neutron
stars, and heavy-ion collisions. Although the motive of FAIR
is to produce baryon-rich QGP, it is uncertain that at what point
on the phase diagram along the baryon density $(\rho_B)$ axis QGP
can be produced.\\
%%%%%%%%%%%%%%%%%%%%%%%%%%%%%%%%%%%%%%%%%%%%%%%%%%%%%%%%%%%%%%%%%%%%%%%%%%%%%%%%%%%%

Since the relative velocity of heavy-quark-bound sate 
($v_{Q \bar Q} \ll 1$) is very small,
in-medium dynamics of $Q \bar Q$ bound states has 
been extensively
studied with  phenomenological potential 
models~\cite{Karsch:ZphysC37/1988}, 
where the temperature-dependent potential carries all medium effects 
with nonperturbative terms borrowed from lattice simulations.
Although the derivation of such models from QCD 
is not established, free energies and other 
quantities~\cite{McLerran:1981pb,Nadkarni:PRD34/1986} 
from lattice calculations obtained using the correlation functions 
of the Polyakov loop are often taken as input for the potential.
These quantities have been thought to be related to the 
color-singlet and color-octet heavy quark potentials at finite 
temperature~\cite{Nadkarni:PRD34/1986,Nadkarni:PRD33/1986} however, 
 a precise answer is still missing~\cite{Philipsen:NPA820/2009}.
It was until recently that the smallness of the velocity 
($v_{Q \bar Q} \sim \alpha_s <1$) has opened up the possibility of studying 
heavy quark bound sate at finite temperature, 
where both the non-relativistic scales, {\em viz.} the heavy quark mass
($m_Q$), the momentum exchange ($m_Q v_Q$), the binding energy 
($m_Q v_Q^2$) etc.  and the thermal scales, {\em viz.} 
$T$, $gT$, $g^2T$ are are hierarchically ordered:
By exploiting the aforesaid separation of 
scales, a sequence of low-energy effective field theories
(EFTs)~\cite{Escobedo:2008sy,Brambilla:2008cx} 
{\em viz.} non-relativistic QCD (NRQCD), potential non-relativistic 
QCD (pNRQCD) {\em etc.} have been synthesized by integrating out the 
successive energy scales. In this context, pNRQCD and its thermal version
are of particular interest because it describes the quarkonium 
dynamics through potentials and low-energy interactions
~\cite{Pineda:1997bj,Brambilla:1999xf}. 
The color-screening phenomenon~\cite{Pineda:1997bj} and 
thermal width are thought to be an outcome of thermal corrections
to the real and imaginary parts of the color-singlet potential,
respectively. Out of various mechanisms that might be 
contributing to the imaginary part of the potential, 
Landau-damping~\cite{Laine:2006ns} is the dominant one, which is 
responsible for quarkonium dissociation at weak coupling.  
It makes the quarkonium to be dissociated even at temperatures,
where the possibility of color screening is negligible. 
Moreover, lattice studies have also shown that even at strong 
coupling~\cite{Laine:2007qy,Rothkopf:2011db,Burnier:2012az} 
the potential  may have a sizeable imaginary part and 
such contributions  are related to quarkonium decay processes 
in the plasma.\\ 
 
The works referred above are restricted to finite temperature 
only. However, the color screening effect at finite temperature 
and finite density has been studied in thermo-field dynamics
approach to calculate the Debye screening mass, $m_{{}D}(T,\mu)$,
where authors  in Refs.~\cite{Gao:PLB378/1996} 
and~\cite{Blin:PRC55/1997} have used the phenomenological
potential model~\cite{Karsch:ZphysC37/1988} and  error 
function-type confined force with color screened 
Coulomb-type potential, respectively.
Applying the effective perturbation theory for gauge theories 
at finite temperature~\cite{Braaten:NPB337/1990}, authors
in Ref.\cite{Hans:PLB342/1995} extended it to the finite chemical potential 
for studying the collisional energy-loss of  heavy quark in QGP.
Photon production at 
finite chemical potential in QCD plasma at leading-order in strong 
coupling has been computed in~\cite{HGervais:PRC86/2012}. 
Furthermore the dissipative hydrodynamic effects on QGP at 
finite density has been studied by developing a causal 
dissipative hydrodynamic model at finite baryon density for RHIC 
and LHC energies to study the net-baryon 
rapidity distribution~\cite{Monnai:jphyconfser432/2013}. 
Although, finite density lattice QCD calculations are seriously
affected by the sign problem, recently color screening in 
heavy quark potential at finite density with Wilson fermions 
has been studied in lattice QCD in both real and imaginary 
chemical potential regions~\cite{Takahashi:PRD88/2013}.\\ 

In the present work we have studied quarkonium dissociation in 
the quark matter of high baryon density and low
temperature. For that we have first obtained the heavy quark
potential by correcting both perturbative and nonperturbative
terms in the Cornell potential through the dielectric function
at finite temperature and chemical potential and then calculate
the binding energy of the quarkonium states. The rest of the
work is organized as follows: We first discuss the heavy quark
potential at finite chemical potential in Sec.II. Then we focus
on the estimation of the binding energy of quarkonium at
finite chemical potential and thereby obtain the dissociation
chemical potential for quarkonium ground states in Sec.III.
Finally we conclude in Sec.IV.

\section{Heavy quark potential at finite chemical potential and running coupling constant}
The interaction potential between  $Q$ and $\bar Q$ is subjected 
to modification when the pair ($Q \bar Q$) is  placed in a hot and dense 
QCD medium, and it plays an important role in understanding the status of 
$Q \bar Q$ bound states in such medium.
This issue has been well taken up by various authors and is
reported in several reviews~\cite{Bram:Revmodphy77/205,kluberg:arxiv/0901.3831},
by dealing on both fronts, the phenomenology~\cite{Karsch:ZphysC37/1988} 
as well as the lattice QCD ~\cite{Laine:2007qy,Rothkopf:2011db,Burnier:2012az,Laermann:PLB173/1998}. 
All these studies assume that the phase transition from a hadronic matter 
to a QGP phase melts the string, and thereby the string tension vanishes 
at the transition point.
However, lattice results~\cite{Karsch:jphysconfser46/2006}
indicate that there is no genuine phase transition at vanishing
baryon density, it is rather a crossover, and 
there may not be immediate melting of string at the 
deconfinement temperature.
Thus one should consider its effect on the potential even above the
transition temperature, in addition to the coulomb term.

The large mass of heavy quark meets the requirements: (i) $m_Q \gg
\Lambda_{QCD}$ and (ii) $T \ll m_Q$ for the  description of the 
interactions between $Q\overline Q$  at finite temperature and
density. It is thus possible to obtain the quantum mechanical
potential V(r,T), by correcting short as well as  long-distance
parts  of the $Q \bar Q$ potential through  dielectric 
permittivity, $\epsilon (k)$, embodying the effects of deconfined medium:
\begin{eqnarray}
\label{defn}
V(r,T,\mu)&=&\int \frac{d^3\mathbf k}{{(2\pi)}^{3/2}}
\left( e^{i\mathbf{k} \cdot \mathbf{r}}-1 \right)~\frac{V(k)}{\epsilon(k)} ~.
\end{eqnarray}
The term  $ V(k)$ in above equation is the Fourier Transform (FT) 
of the potential. However, obtaining 
the FT of the linear term is a tricky part and needs regulation, which 
is carried out by multiplying an 
exponential damping factor and is then switched
off after the evaluation of FT. We have regulated both terms by the same 
screening scale\footnote
{However, in the framework of classical Debye-H\"{u}ckel theory,
different screening functions were employed, {\em viz.}
$f_c$ and $f_s$ for the Coulomb and string terms, respectively 
to obtain the free energy~\cite{Digal:EPJC43/2005} whereas 
in Ref.~\cite{Megias:PRD75/2007},
different scales for the Coulomb and linear pieces have been 
employed through a dimension-two gluon condensate.} and the FT of
the Cornell  potential is thus obtained as~\cite{Vineet:PRC80/2009}
\begin{equation}
\label{ftcornell}
{\bf V}(k)=-\frac{\sqrt{2/\pi}}{k^2}\left(\alpha+\frac{2 \sigma}{k^2}\right),
\end{equation}
where $\alpha$ and $\sigma$ are the coupling and string tension,
respectively. We have taken $\alpha$ as function of chemical
potential ($\mu_q$) and temperature up to two loops ~\cite{PKShrivastava:PRD82/2010} as
\begin{eqnarray}
%\label{alpha1}
%\alpha(T,\mu_q)&=&\frac{12 \pi}{(33-2 N_f)}\left[ ln\left(\frac{0.8\mu_q^2+15.662T^2}{\Lambda^2_T}\right)\right]^{-1},\\
\alpha(T,\mu_q)&=&\frac{6 \pi}{(33-2 N_f)~ln\sqrt{\frac{T^2}{\Lambda^2_T}
+\frac{\mu^2_q}{\pi^2\Lambda_T^2}}} \left[1-\frac{3(153-19N_f)~
ln\left(2~ln\sqrt{\frac{T^2}{\Lambda^2_T}+
\frac{\mu^2_q}{\pi^2\Lambda_T^2}} \right)}{(33-2 N_f)^2~
ln\sqrt{\frac{T^2}{\Lambda^2_T}+\frac{\mu^2_q}{\pi^2\Lambda_T^2}}}\right],
\label{alpha2}
\end{eqnarray}
where $\Lambda_T$ is  the QCD scale-fixing parameter which
characterizes  the strength of the interaction.
% and we have taken $\Lambda_T$= 115 MeV. 
To obtain the dielectric permittivity, $\epsilon(k)$, one has 
to evaluate
the self-energies and the corresponding 
static propagators in weak coupling Hard Thermal 
Loop (HTL) approximation.
In the real-time formalism using Keldysh representation, the 
retarded (R),
advanced (A) and symmetric (F) propagators are written as the linear
combination of the components of the $(2 \times 2)$ 
matrix propagator:
\begin{eqnarray}
\label{2a6}
   D_R^0 = D_{11}^0 - D_{12}^0 ~,~ D_A^0 = D_{11}^0 - D_{21}^0 ~,~
   D_F^0 = D_{11}^0 + D_{22}^0  ~,
\end{eqnarray}
where only the symmetric component involves the distribution
functions and is of particular advantage for the HTL diagrams, 
where the terms containing distribution functions dominate. 
Similar relations hold good for 
the retarded ($\Pi_R$), advanced ($\Pi_A$) and symmetric ($\Pi_F$) 
self energies. The resummation of the propagators can be 
obtained using the Dyson-Schwinger equation
\begin{eqnarray}
 {D}_{R,A}&=&D_{R,A}^0+D_{R,A}^0\Pi_{R,A}{D}_{R,A}~, \label{2b2}\\
 {D}_{F}&=&D_{F}^0+D_{R}^0\Pi _R{D}_{F}+D_F^0\Pi_{A} {D}_{A}+ 
 D_{R}^0\Pi _{F}{D}_{A}~. 
\label{2b7}
\end{eqnarray}
For the static potential, we need only the temporal component
(``00" $\equiv $ L) of the propagator, whose evaluation is
easier in the Coulomb gauge. Thus the above resummation (\ref{2b2})
can be recast through its temporal component as
\begin{eqnarray}
 D^L_{R,A}=D^{L(0)}_{R,A}+D^{L(0)}_{R,A}\Pi^L_{R,A}{D}^L_{R,A}~. \label{3b2}
\end{eqnarray}
The leading contribution to  temporal component of the
retarded, advanced and symmetric  gluon self-energy in the 
HTL-approximation can be written as 
\begin{eqnarray}
\Pi^{L}_{R,A}(k)&=& m_{{}D}^2\left(\frac{k_{0}}{2k}\ln\frac{k_{0}+k\pm i\epsilon}{k_{0}-k\pm i\epsilon}-1\right)
\end{eqnarray}
and
\begin{eqnarray}
\label{imself}
\Pi^{L}_{F}(k)&=&-2\pi i m_{{}D}^2\frac{T}{k}\Theta(k^2-{k_0}^2)~,
\end{eqnarray}
respectively. Thus the gluon self-energy is composed of real and imaginary 
parts which are responsible for the Debye screening and the 
Landau damping,
respectively. The real part of potential can be obtained from 
the retarded (or advanced) self energy with the prescriptions
$+i\epsilon $ ($ -i\epsilon $), respectively.\\
The term ${m_{{}_D}}$ in above equation is known as electric screening 
mass (Debye screening mass) which encodes the medium effects in terms 
of temperature and baryon chemical potential.
The properties of QCD at finite temperature and density are 
studied in~\cite{Kajantie:AnnPhys160/1985}  and the 
electric screening mass can be obtained in the static infra-red 
limit of ``00" component of gluon self energy 
$\Pi^{00}(q_0=0, \vec q \rightarrow 0)$. Taking the dynamical 
quark mass  $m_q$=0, the electric screening  mass at 
finite temperature and  vanishing baryon chemical
potential ($\mu=\mu_b=0$) based on perturbation 
theory in high temperature limit is
~\cite{Kalashnikov:Fortschrphy32/1984}, 
%function, $ \epsilon(k)$~\cite{prc_vineet} 
\begin{equation}
\centering
\Pi^{00}(q_0=0, \vec q \rightarrow 0)=\left(\frac{N_c}{3}+\frac{N_f}{6}
\right) gT,
\end{equation}
where $N_c$ and $N_f$ are number of colors  and flavours of 
color group SU($N_c$) respectively. For a hot and dense plasma, 
the resummation technique to the perturbation theory proposed
by~\cite{Braaten:NPB337/1990} works well in momentum scale $gT$. 
The leading-order Debye screening mass 
from HTL~\cite{Hans:PLB342/1995,Ebraaten:PRD42/1990, 
RDPisarski:NPA525/1991,RDPisarski:PRD47/1993} at finite 
temperature and chemical  potential is given by
\begin{equation}
{m^2_{D}(T,\mu_q)}= g^2(T) T^2 \left(\frac{N_c}{3}+\frac{N_f}{6}+ 
\frac{1}{2 \pi^2}\sum_f \frac{\mu_q^2}{ T^2}\right),
\label{debyechem}
\end{equation}
where the quark chemical potential is related to the baryon chemical
potential by $\mu_q=\mu_b/3$. Later the screening mass at finite 
temperature and baryon density has also been studied on the lattice 
through the Taylor expansion method~\cite{Doring:POSLAT2005/2006}, 
which coincides with the above equation. For zero 
chemical potential ($\mu_q$=0), the leading-order Debye mass~(\ref{debyechem})
reduces to
\begin{equation}
 m_{{}D}(T)= g(T)T\sqrt{\frac{N_c}{3}+\frac{N_f}{6}}.
 \label{debyemu0}
\end{equation}

The next-to-leading contribution to Debye mass 
($\delta{m_{D}}(T,\mu_q)$) 
comes from the resumed one-loop gluon diagrams
~\cite{Braaten:NPB337/1990}.
However, we restrict ourselves only up to the leading-order term in the
Debye mass. The variations of the Debye mass in leading-order 
with the chemical potential and temperature are shown in Fig. 1. 
The left panel of Fig.1 
shows that the Debye mass increases with the chemical potential 
and is more for higher temperature. This implies that the
 contribution to the Debye mass at low temperature and high 
baryon chemical potential mainly comes 
from the chemical potential, which can be seen in the right panel of Fig.1.
For low temperature region (20-50 MeV) the 
increase in Debye mass is negligible but increases with $\mu_b$.
The temporal component of the retarded (or advanced)  propagator 
in the static limit~\cite{Latathakur:PRD89/2014} is %({\bf HOW do you get this expression?}) :
\begin{eqnarray}
 D^{00}_{R,A}(0,k)=-\frac{1}{(k^2+m_{{}D}^2(T,\mu))},
\label{rtrdprop}
\end{eqnarray}
Hence we can now obtain the dielectric permittivity from the 
static limit of the ``00"-component of gluon propagator 
\begin{eqnarray}
\label{dielec}
\epsilon^{{}^{-1}}(k)&=&-\lim_{\omega \to 0} {k^2} D_{11}^{00}(\omega, k)~~;~~
D^{00}_{11}(\omega,k)=\frac{1}{2}\left( D^{00}_{R}+D^{00}_{A}\right)
\end{eqnarray}
Thus the medium modified potential is  obtained by substituting
Eqs. (\ref{ftcornell}) and (\ref{dielec}) into Eq.(\ref{defn}) 
\begin{eqnarray}
V({\bf r},T,\mu)&=&\int \frac{d^3\mathbf k}{{(2\pi)}^{3/2}} 
 (e^{i\mathbf{k} \cdot \mathbf{r}}-1)  \left(-\frac{\sqrt{2/\pi}}{k^2}\left(\alpha+\frac{2 \sigma}{k^2}\right)   \right) \left[\frac{k^2}{(k^2+m_{{}D}^2(T,\mu))}\right]
\end{eqnarray}
\vspace{1.0in}
% % % % % % % % % % % % % % Figure 1 % % % % % % % % % % % % % % % % % % % %
\begin{figure}[h!]
\label{fig1}
\begin{center}
\includegraphics[scale=0.45]{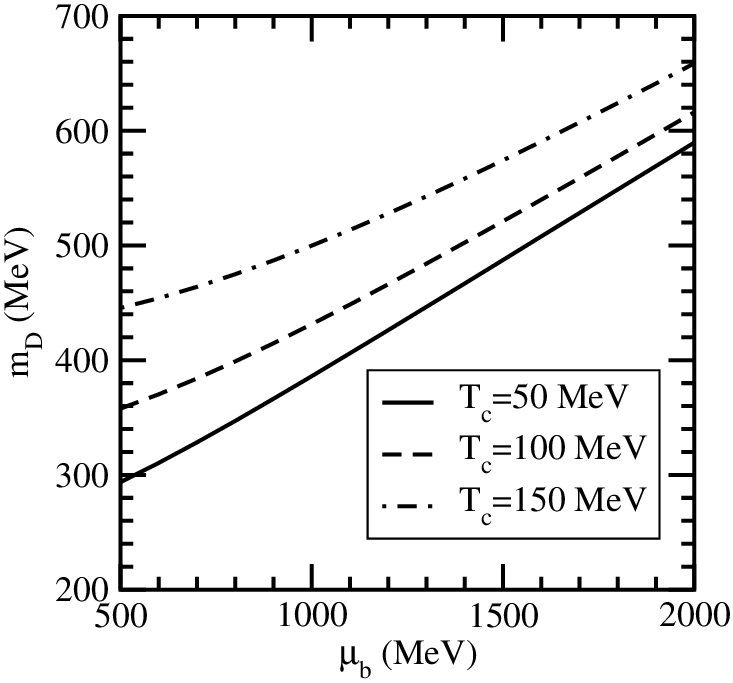}
\hspace{0.3in}
\includegraphics[scale=0.45]{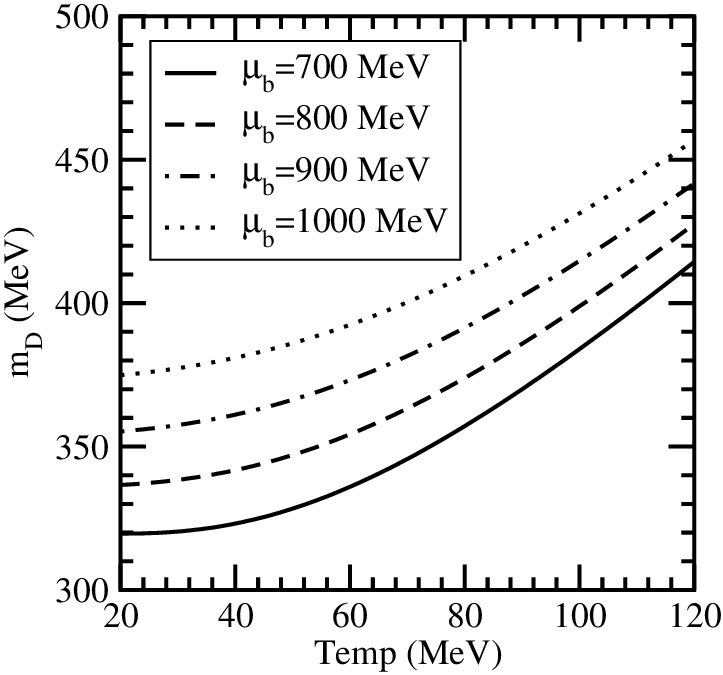}
\caption{Variation of the screening mass ($m_{{}D}$) against the baryon  
chemical potential ($\mu_b$)  at different values of 
temperature (left panel). Same against the temperature at
different values of baryon chemical potential (right panel).}
\end{center}
\end{figure}
Solving the integral in above equation,  one can to obtain 
the  medium 
modified potential at finite density and temperature 
(with $\hat r = r m_{{}D}(T,\mu)$) as:
\begin{eqnarray}
\label{hqpot}
{\bf V}(r,T,\mu)&=&\frac{2\sigma}{m_{{}_D}(T,\mu)}\left(\frac{e^{-\hat r}-1}{\hat r} +1 \right)-\alpha
m_{{}D}(T,\mu)\left(\frac{e^{-\hat r}}{\hat r} +1 \right),
\end{eqnarray}
where the chemical potential and temperature dependencies 
are introduced through the Debye mass, $m_{{}D}(T,\mu)$. 
The potential has a long range 
Coulombic tail in addition to the standard Yukawa term and 
the constant terms are introduced  to yield the correct limit 
of $V(r,T,\mu)$ as $T\rightarrow 0,\mu\rightarrow 0$  
(it reduces to the Cornell potential).\\

Recently it is known that there exists imaginary part of the potential
~\cite{Escobedo:2008sy,Brambilla:2008cx,Laine:2007qy,Beraudo:NPA806/2008}  
which  originates from the static limit of symmetric self energy 
(\ref{imself}) and plays an important role in weakening the bound state 
peak that leads to a finite width ($\Gamma$) for the resonance peak in 
the spectral function. The  imaginary part of the temporal component of 
symmetric propagator gives the corresponding  dielectric function~\cite{Latathakur:PRD89/2014}
\begin{eqnarray}
\label{imsympropagator}
\epsilon ^{-1}(0,k)= \frac{-\pi T m_D^2(T,\mu) k^2 }{k (k^2+m_{{}D}^2(T,\mu))^2}~,
\end{eqnarray}
which, in turn, gives the imaginary part of $Q\bar Q$ potential:  
 \begin{eqnarray}
{\Im V}({\bf r},T,\mu)&=&-\int \frac{d^3\mathbf{k}}{(2\pi)^{3/2}}
(e^{i\mathbf{k} \cdot \mathbf{r}}-1)
\left(-\frac{\sqrt{2/\pi}}{k^2}\left(\alpha+\frac{2 \sigma}{k^2}\right)\right)
k^2\left[\frac{-\pi T m_D^2(T,\mu)}{k(k^2+m_D^2(T,\mu))^2}\right]\nonumber\\
&=& -\alpha T\phi_0( \hat r)+ \frac{2\sigma T}{m_D^2(T,\mu)}\psi_0(\hat{r})~,
\end{eqnarray}
where the first (second) term comes from contribution due to the 
Coulomb (linear) term in the 
leading-order~\cite{Dumitru:PRD792009,Dumitru:PRD79540192009,Margotta:2011ta} 
with the following functions, $\phi_0(\hat r)$ and $\psi_0(\hat r)$, 
respectively:
\begin{eqnarray}
 \phi_0(\hat r)&=&-\frac{{\hat{r}}^2}{9}
\left(-4+3\gamma_{E}+3\log\hat{r}\right)\\
\psi_0(\hat{r})&=&\frac{\hat r^2}{6}+\frac{-107+60\gamma_E
+60\log(\hat r)}{3600}\hat r^4+O(\hat r^5)~.
\end{eqnarray}
Thus in leading-logarithmic order, the imaginary component of potential 
becomes
\begin{eqnarray}
\label{fullimgpot}
\Im V{(r,T,\mu)}&=&-T\left(\frac{\alpha {\hat r^2}}{3}
+\frac{\sigma {\hat r}^4}{30m_D^2(T,\mu)}\right)\log(\frac{1}{\hat r})~,
\end{eqnarray}
which shows that imaginary part vanishes for small distances~\cite{Latathakur:PRD89/2014}.
However, its magnitude is larger than the case 
where only the Coulombic term is 
considered~\cite{Dumitru:PRD792009,Dumitru:PLB6622008,Burnier:2009yu}, due
to the additional linear (string) term.\\
Now with the imaginary potential (\ref{fullimgpot}), the width for a 
particular resonance state can be calculated as: 
\begin{eqnarray}
\label{gammaa}
\Gamma(T,\mu)/2 &=&\int d^3 {\bf r}|\Psi(r)|^2    \left[\frac{\alpha T{\hat r^2}
}{3}
+\frac{T\sigma {\hat r^4}}{30 m_D^2(T,\mu)}
\right]\log(1/\hat r)\nonumber\\
&=&T\left(\frac{4}{\alpha m_Q^2}+\frac{12\sigma}{\alpha^2m_Q^4}\right)
m_D^2 \log\frac{\alpha m_Q}{2m_D(T,\mu)}.
\end{eqnarray}
Here for simplicity,  we have taken $\Psi(r)$ as the 
Coulombic wave function, similar to the ground state
wave function in hydrogen atom.
This shows that $\Gamma$ is function of both chemical 
potential and temperature through the screening mass (Debye mass).
However, the dependences on temperature and chemical potential 
are opposite, {\em namely} 
$\Gamma$ increases with temperature but decreases with chemical 
potential. That is why $\Gamma$ does not have much role in dissociation
of quarkonia states unlike at high temperature. This will be discussed 
in the next section.\\
%%%%%%%%%%%%%%%%%%%%%%%%%%%%%%%%%%%%%%%%%%%%%%%%%%%%%%%%%%%%%%%%%%%%%%%%%%%%%%%%
It is worth to note that the potential
in a hot QCD medium is not the same as the lattice parametrized 
heavy quark free-energy in the deconfined phase which is 
basically a screened Coulomb~\cite{HSatz:RPP63/2000} 
because one-dimensional Fourier transform of the Cornell potential 
in the medium yields the similar form as used in the lattice 
QCD to study the quarkonium properties which assumes the 
one-dimensional color flux tube structure~\cite{dixit:MPLA5/1990}.
However, at finite temperature and density that may not be
the case since the flux tube structure may expand in more
dimensions~\cite{HSatz:RPP63/2000}, hence the three-dimensional 
form of the  medium modified Cornell potential may therefore be the 
better option.
The potential (\ref{hqpot}) leads to an analytically solvable 
Coulomb potential if one neglect Yukawa term in the 
limit $r>>1/m_{{}D}$ and the product
$\alpha m_{{}D}$ will be much greater than $2\sigma/m_{{}D}$ 
for large values of chemical potentials. 
%({\bf But here is the large chemical potential, not large temperature?}). 
\begin{eqnarray}
\label{lrp}
{V(r,T,\mu)}\sim -\frac{2\sigma}{m^2_D(T,\mu)r}-\alpha m_{{}D}(T,\mu)
\end{eqnarray}
To see the effects of a hot and dense medium on the $Q \bar Q$ 
potential, we have evaluated the potential at large chemical potential, 
{\em viz.} at 600 MeV, 800 MeV and 1000 MeV in the low 
temperature range 20-60 MeV in Fig.2.
We have found that the potential rises with the interquark distance, $r$, 
but slower than linearly, {\em i.e.} long-range QCD force becomes 
short-range and at some large enough $r$, it simply flattens
out due to the breaking of QCD string, {\em i.e.} a heavy meson 
splits into two heavy-light mesons. This observation agrees with 
the lattice  calculations~\cite{OKaczmarec:hep-lat/0312015,Burnier:PRL114/2015}.
Thus, the deconfinement is reflected in the
large-distance behaviour of heavy quark potential
even at large chemical potential. That is why
the in-medium behaviour of heavy
quark bound states is used to probe the state of matter in QCD 
thermodynamics at finite density and/or temperature. 
% % % % % % % % % % % % % % % % Figure 2 % % % % % % % % % % % % % % % % % %
%\vspace{1.5in}
\begin{figure}
\begin{center}
\includegraphics[scale=0.5]{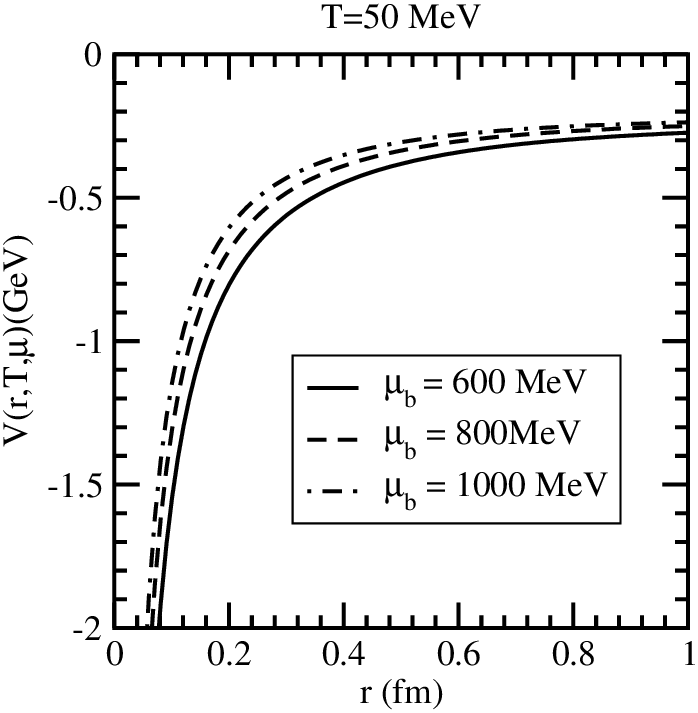}
\hspace{0.3in}
\includegraphics[scale=0.5]{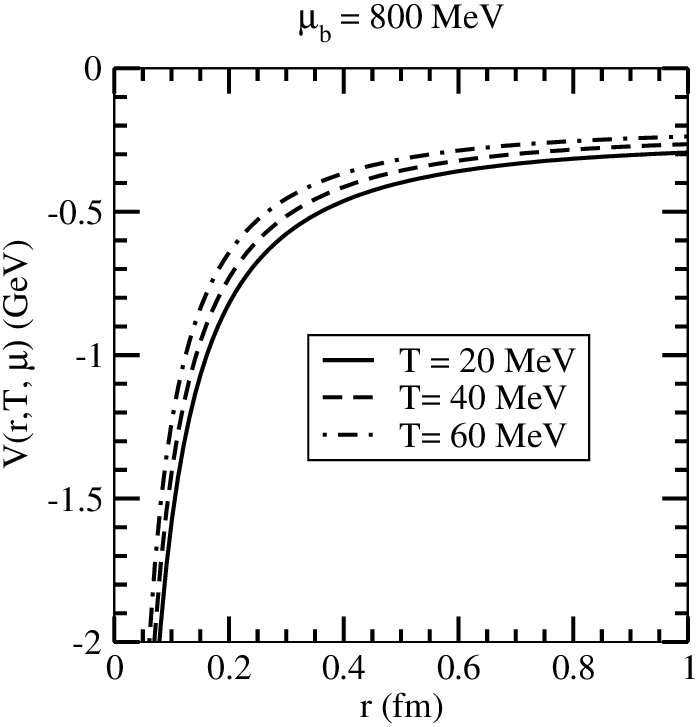}
\caption{ 
Variation of medium modified   $Q\overline Q$ potential in a hot
and dense QCD medium at given low temperature with different 
baryon chemical potential $\mu_b$ (left panel). Same for different temperatures at given  $\mu_b$ (right panel).}
\label{pot_vs_chempot}
\end{center}
\end{figure}
%%%%%%%%%%%%%%%%%%%%%%%%%%%%%%%%%%%%%%%%%%%%%%%%%%%%%%%%%%%%%%%%%%%%%%%%%%%%%%%%%%%%%
\section{Binding energy and dissociation chemical potential}
The  medium-modified effective potential  obtained in the
previous section has been employed to study the in-medium 
properties of the heavy quark  bound states such as, binding 
energies and  dissociation chemical potentials.
We studied the binding energy and the dissociation chemical potential 
for  the ground states of  $c \bar c$. 
In order to understand bound state properties of quarkonium 
states in hot and dense QCD medium, one need to solve the
 Schr\"{o}dinger equation using
the $Q\overline Q$ potential. It can be seen from   
Eq.(\ref{hqpot}) that  in the short-distance limit, the 
vacuum contribution of $Q\overline Q$
potential supersedes over the in-medium contribution whereas 
the in-medium contribution affects the potential in the 
long-distance limit.
As we have already seen the potential (\ref{lrp}) in long-distance
($rm_D \gg 1$) limit and for large values of chemical potential (or 
temperature) is Coulomb-like potential after identifying
$2\sigma/m_{{}D}^2$
with the square of color charge, $g_s^2$, so the energy eigenvalues
are read as in hydrogen atom problem:
\begin{eqnarray}
E_{\rm{bin}}\stackrel{\hat{r}\gg 1}{=}\left( \frac{\sigma m_Q }{m_{{}_D}^2(T,\mu) n^{2}} +\alpha 
m_{{}_D}(T,\mu) \right);~n=1,2 \cdot \cdot \cdot
\end{eqnarray}

However, in the intermediate-distance ($rm_D(T,\mu) \simeq 1$) 
scale, the
interaction becomes non-trivial and the potential
does not look simpler in contrast to the asymptotic limits, thus
the full potential (16) in general needs to be dealt numerically to obtain 
the  binding energies. 
The matrix method serves the purpose and the stationary
Schr\"odinger equation can be solved in a matrix form 
through a discrete basis, instead of the continuous real-space 
position basis spanned by the states $|\overrightarrow{x}\rangle$. 
Subdividing the potential $V$ (16)
into N discrete wells having potentials $V_{1}$ through $V_{N+2}$,
 such that
$V=V_{i}$ for $x_{i-1} < x < x_{i};~i=2, 3,...,(N+1)$ 
is the $i^{\rm{th}}$
boundary potential. For existence for a bound state, there must 
be an exponentially decaying wave function  in the region $x > x_{N+1}$ 
as $x \rightarrow  \infty $ and the wave function in this 
region reads:
\begin{equation}
\Psi_{N+2}(x)=P_{{}_E} \exp[-\gamma_{{}_{N+2}}(x-x_{N+1})]+
Q_{{}_E} \exp [\gamma_{{}_{N+2}}(x-x_{N+1})] ,
\label{shcrod}
\end{equation}
where, $P_{{}_E}= \frac{1}{2}(A_{N+2}- B_{N+2})$,
$Q_{{}_E}= \frac{1}{2}(A_{N+2}+ B_{N+2}) $ and,
$ \gamma_{{}_{N+2}} = \sqrt{2 \mu_r(V_{N+2}-E)}$. The energy eigenvalues 
can then be identified with the zero's of $Q_E$.

\begin{figure}
%\vspace{0.2in}
\begin{center}
\includegraphics[scale=0.45]{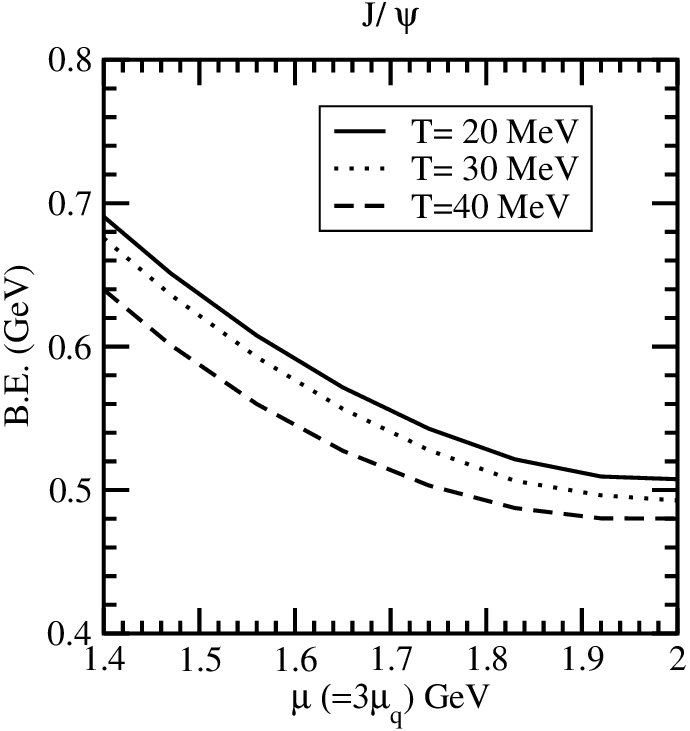}
%\hspace{0.3in}
%\includegraphics[scale=0.45]{gama_vs_mu.eps}
\caption{Variation of binding energy of $J/\psi$ state
in a hot/dense QCD medium with baryon chemical potential at 
different values of temperature.}
\label{be}
\end{center}
\end{figure}

Thus, matrix method gives the binding energies as a function of chemical 
potential in a small 
temperature range (20 - 40 MeV), which are shown in Fig. \ref{be}. It is found
that the binding energy decreases with the increase of chemical 
potential. This can be understood by the fact that the screening gets 
stronger with the increase of the baryon chemical potential and
hence the inter-quark potential becomes weaker compared 
to $\mu=0$. Thus the study of the binding energies at finite temperature and
chemical potential provides a  information about 
the dissociation pattern of quarkonium states in the region of 
high baryon chemical potential and low temperature.\\
%%%%%%%%%%%%%%%%%%%%%%%%%% Figure 3 %%%%%%%%%%%%%%%%%%%%%%%%%%%%%%%%
Now one can obtain the dissociation chemical potential
of quarkonia when their binding energies in the dense baryonic medium 
are of the order of the baryon chemical potential.
We found that with the rise in temperature, the 
dissociation chemical 
potential deceases (Table 1) slightly, {\em for example,} at 
temperature $T$ = 20 MeV, the $J/\psi$
will be dissolved at 1.42 $\mu_c$ whereas for T=30 Mev, $\mu_D$ becomes 
1.40 $\mu_c$\footnote{{The values of $T_c$ and $\mu_c$ used in this work are taken from the Ref.~\cite{Jkapusta:PRC81/2010}, {\em viz.} the critical
baryon chemical potentials ($\mu_c$) are 1189 MeV, 1183 MeV, 1169 MeV and 
1154 MeV for the critical temperatures ($T_c$) 20 MeV, 30 MeV, 40 MeV and 50 MeV,  respectively.}} etc.  
This observation can be used to locate the point on QCD phase diagram
in the high-baryon density limit 
where the confined nuclear matter cross to the phase boundary and
becomes baryon-rich deconfined quark matter
% % % % % % % % % % % % % % % % % Table 1 % % % % % % % % % % % % % % % % % %
\begin{table}
\label{table1}
\centering
\begin{tabular}{|c|c|c|c|}
\hline
{$T$ (MeV)} &$\mu_D$ (MeV) & $\mu_D/\mu_c$  \\
\hline
 20 & 1683 & 1.42   \\
\hline
30 &1659&1.40\\ 
\hline
40 &1611&1.38\\ 
\hline
50 &1560&1.35\\ 
\hline
\end{tabular}
%\vspace{0.7mm}
\caption{Dissociation chemical potential for $J/\psi$ at 
different temperatures.}
\end{table}
Recently the physics of dissociation has been further complicated by
arguing that the potential should have an imaginary part
as well, which, like a width, facilitates the dissociation of quarkonia
further. Thus, nowadays the dissociation temperatures are obtained from
the width, $\Gamma$, where the dissociation is expected to occur 
when binding energy decreases with the chemical potential 
and becomes  $\sim \Gamma/2$ ~\cite{Mocsy:PRL99/2007,Burnier:JHEP01/2008}.
We have found that the dissociation chemical potential estimated 
from the above criterion is almost same to the values obtained 
from the binding energy criterion (Table 1).

\section{Conclusion}
We have studied the quarkonium dissociation in hot and 
dense QCD medium by correcting both the perturbative and 
nonperturbative terms of the Cornell potential through the
dielectric permittivity Thereafter the chemical potential-
and temperature-dependent potential is plugged into
Schr\"{o}dinger equation to study the properties of quarkonia states
in high baryon density and low temperature region.

It is noticed that the screening mass increases dominantly with the chemical 
potential rather than with the temperature in the high baryon density 
and low temperature region. As a consequence the binding energies 
decrease with the chemical potential. The dissociation chemical 
potentials of $J/\psi$ at $T$= 40 and 50 MeV is found to be 1611 MeV and 
1560 MeV, respectively. Finally, we conclude that the dissociation 
of quarkonia states in large baryon density may help to trace out 
the point on the QCD phase diagram at which the baryon 
rich QGP (which is expected at FAIR energies) will be created.\\

\noindent {\bf Acknowledgement:}
We are thankful to Lata Thakur for her sincere help in deriving the
quarkonium potential. B.K.P.  is thankful to Government of India for 
financial assistance under CSIR project (CSR-656-PHY). 
Uttam Kakade is thankful to the Government of Maharashtra for availing study leave.\\
%%%%%%%%%%%%%%%%%%%%%%%%APPENDIX%%%%%%%%%%%%%%%%%%%%%%%%%%%%%%%%%%%%%%%%%%%%
%%%%%%%%%%%%%%%%%%%%%%%%%%%%%%%%%%%%%%%%%%%%%%%%%%%%%%%%%%%%%%%%%%%%%%%%%%%%%%%%%%%%%%%%%%%%%%%%%%%%%%%%%%%%%%%%

\end{document}